\documentclass{optica-article}

\journal{opticajournal} 

\articletype{Research Article}

\usepackage{lineno}
\usepackage{booktabs}
\usepackage{subfigure}

\begin{document}

\title{Reconfigurable structural color generation in multicolor pixels using phase-change materials}

\author{Md Tanvir Emrose,\authormark{1,2}
Katherine Winchester,\authormark{1}
Yin Huang,\authormark{3,4}
and Georgios Veronis\authormark{1,2,*}}

\address{\authormark{1}School of Electrical Engineering and Computer Science, Louisiana State University, Baton Rouge, Louisiana 70803, USA\\
\authormark{2}Center for Computation and Technology, Louisiana State University, Baton Rouge, Louisiana 70803, USA\\
\authormark{3}School of Physics, Central South University, Changsha, 410083, China\\
}

\email{\authormark{4}yhuan15@csu.edu.cn\\
\authormark{*}gveronis@lsu.edu} 


\begin{abstract*} 
We introduce multilayer structures based on phase-change materials for reconfigurable structural color generation. These structures can produce multiple distinct colors within a single pixel. Specifically, we design structures that generate either two or four maximally distinct structural colors. We employ a memetic optimization algorithm coupled with the impedance method to identify the optimal combination of materials and layer thicknesses that maximizes the distinctiveness of the generated colors. We demonstrate that our design approach achieves strong color contrast between the generated colors. The proposed multilayer design eliminates the need for subwavelength lithography, making it highly suitable for large-scale applications. Our results could lead to a new class of single-cell multicolor pixels which retain each color without power consumption, making them particularly appealing for low refresh rate displays.
\end{abstract*}

\section{Introduction}

Structural color generation, arising from the interaction between light and nanostructures, offers a promising alternative to traditional dye- and pigment-based methods, with several significant advantages. Conventional organic dyes and chemical pigments are susceptible to chemical instability and degradation under heat or ultraviolet radiation, limiting the longevity and stability of the colors they produce \cite{wang2023_3}. In addition, their limited spatial resolution, reduced spectral purity, and potential environmental impact make them less suitable for high-resolution color displays and spectral imaging \cite{wang2023_3}. In contrast, structural coloration provides high spatial resolution, long-term stability, and environmentally friendly fabrication processes \cite{wang2023_3}. These advantages are evident in examples in nature such as the iridescent feathers of peacocks and the scales of butterflies \cite{ghiradella1999_3_17, ghiradella1991_3_18}, as well as in ancient artifacts such as stained glass and the Lycurgus Cup \cite{barber1990_3_20, hunault2017_3_22}. Recent progress in computational electromagnetics and nanofabrication techniques has enabled the development of highly versatile structural color devices, with applications ranging from imaging and displays to sensing and energy harvesting \cite{wang2023_3}.
Some of these devices are based on plasmonic structures \cite{lee2018_7_9}, photonic crystals \cite{wu2020_7_11}, or thin films \cite{kats2013_5_8} to produce vivid, passive colors. The resulting optical properties are determined by the geometric configuration and material composition of the nanostructures, which determine the observed color \cite{kim2021_7}. Recent advancements have focused on expanding the achievable color gamut \cite{flauraud2017_7_13, wang2017_7_14}, enhancing brightness \cite{yang2020_7_17}, minimizing pixel size \cite{williams2017_7_16}, and achieving angle-independent coloration \cite{chung2012_7_19}. Additionally, multi-color generation is being explored through the manipulation of thin-film layer thicknesses and the refractive index of materials in stacked structures \cite{ji2016_5_9, lyu2017_5_10}. By refining these methods, structural color filtering has emerged as a highly promising, chemical-free approach to coloration, offering a broad color gamut, wide viewing angles, and scalable fabrication techniques \cite{santos2021_4}. These attributes support its potential use in diverse fields, including anti-counterfeiting technologies \cite{meng2018_7_4, nam2016_7_5}, laser printing \cite{wei2019_5}, novel color displays \cite{franklin2017_7_2}, and other emerging applications \cite{xu2011_3_23, yang2019_3_24, wu2021_3_25, kristensen2016_3_27}.

Tunable color generation, especially in display systems and optical components \cite{jafari_2019_1}, plays a critical role in a wide range of applications, including smart glass, security marking, and active photonics \cite{rios2016_9, macleod2010_9_1}. This technology benefits from advances in optical coatings, multilayer architectures, and structured films, which enable a broad palette of colors in both reflective and transmissive devices. Conventional active display systems, such as those based on light-emitting diodes (LEDs), are capable of producing vivid colors with improved quality and reduced cost \cite{sekitani2009_1_6, cho2015_1_7}. However, these systems require a continuous power supply to sustain color, resulting in significant energy consumption \cite{jafari_2019_1}. In contrast, dynamic color generation offers high resolution, broad color tunability, and enhanced energy efficiency by integrating functional materials within nanophotonic and plasmonic structures without the need for a continuous power source. Instead, it relies on external stimuli to induce reversible changes in optical properties. This approach enables the development of versatile, tunable color filters well-suited for modern applications \cite{sreekanth2021_8, xiong2019_8_2}.

Various functional materials have been investigated for tunable color generation, including liquid crystals \cite{sharma2020_8_9}, electrochromic polymers \cite{xu2016_8_10}, and phase-change materials (PCMs) \cite{shu2018_8_12, wuttig2017_8_13}. Among these, PCMs based on chalcogenide semiconductor alloys have attracted significant interest due to their large refractive index change when switching between crystalline and amorphous phases \cite{sreekanth2021_8}. This switching process is nonvolatile, fast, reliable, and repeatable, making PCMs ideal for use in color filters with nonvolatile pixels capable of retaining their color even after the external stimulus is removed \cite{sreekanth2021_8}. This unique combination of properties enables PCMs such as Ge$_2$Sb$_2$Te$_5$ (GST) to exhibit distinct optical states, resulting from changes in both the real and imaginary parts of the refractive index during phase transitions, thereby enabling effective color switching \cite{hosseini2014_9_19, schlich2015_9_25}.

In this paper, we design multilayer structures incorporating PCMs to generate maximally distinct structural colors. We use a memetic optimization algorithm to identify the optimal combination of materials and layer thicknesses that maximizes the distinctiveness of the generated colors. For two maximally distinct colors, the optimized seven-layer design comprises two thin GST films separated by two dielectric spacer layers, Al$_2$O$_3$ and Ta$_2$O$_5$. It also includes two additional dielectric layers (SiO$_2$ and SiC) above the first GST layer, as well as a MgF$_2$ layer between the second GST layer and the silver substrate.
To achieve two distinct colors, we consider the structure in two different states, with the GST layers either in the crystalline or amorphous phase. Both resulting colors lie near the edge of the chromaticity diagram, leading to a large color separation of $\sim$0.54. Although the structure is designed for normally incident light, it retains its distinctive color-generation capability at viewing angles up to $\sim15^\circ$. 
We also apply the memetic optimization algorithm to design multilayer structures capable of producing four maximally distinct colors. In this case also, the optimized eight-layer design includes two thin GST films. The top GST film is located beneath layers of MgF$_2$ and SiC, while the bottom film is situated above layers of MgF$_2$ and SiC on top of the silver substrate. The two GST layers are separated by dielectric layers of HfO$_2$ and Al$_2$O$_3$. To obtain four distinct colors, we consider all four possible phase combinations of the two GST layers. The minimum distance between any two color coordinates in the optimized structure is $\sim$0.27, and its color contrast performance surpasses that of previously reported designs. Similar to the two-color structures, this design also exhibits stable color generation for incident angles up to $\sim15^\circ$. 
The proposed multilayer structures eliminate the need for subwavelength lithography, making them well-suited for large-scale applications.

The remainder of this paper is organized as follows. In Section \ref{Theory_2} we introduce structural color generation and the merit functions for two- and four-color structures based on the two phases of GST. We also describe the memetic algorithm employed for the optimization of the structures. In Sections \ref{2color} and \ref{4color}, we apply the memetic optimization algorithm to design structures generating two and four maximally distinct colors, respectively. Finally, Section \ref{Conclusion_2} summarizes our conclusions. 

\section{Theory}
\label{Theory_2}

Color perception is a subjective experience influenced by the interaction between electromagnetic radiation and the human visual system \cite{byrne2003_3_40,shevell2003_3_41,conway2010_3_42}. When electromagnetic radiation from a light source illuminates an object, it undergoes reflection, transmission, and absorption. The light that is reflected or transmitted is then captured by the photoreceptors in the human retina, which send signals to the visual cortex, where color perception occurs. This perception is shaped by the optical spectra of the light source, the optical properties of the object, and the physiological response functions of the observer \cite{wang2023_3}.

To quantitatively relate visible spectra to perceived colors, various color spaces have been developed \cite{gowda2019_3_49, ramanath2004_3_48, bora2015_3_47}. Among these, the CIE 1931 color spaces, including XYZ, were the first to establish a quantitative link between the visible spectrum and the colors physiologically perceived by humans. The CIE 1931 XYZ color space was derived from extensive color-matching experiments designed to capture the full range of human color perception \cite{son2011_3_57, schanda2007_3_58}. It defines colors through the tristimulus values $X$, $Y$, and $Z$, which are computed by
\begin{equation}
\left\{
\begin{aligned}
X &= \frac{1}{k} \int_{\lambda_1}^{\lambda_2} \overline{x}(\lambda) I(\lambda) R(\lambda, \theta) \, d\lambda, \\
Y &= \frac{1}{k} \int_{\lambda_1}^{\lambda_2} \overline{y}(\lambda) I(\lambda) R(\lambda, \theta) \, d\lambda, \\
Z &= \frac{1}{k} \int_{\lambda_1}^{\lambda_2} \overline{z}(\lambda) I(\lambda) R(\lambda, \theta) \, d\lambda,
\end{aligned}
\right.
\label{eq_2.1}
\end{equation}
where ($\lambda_1, \lambda_2$) is the spectral range considered, 
$\overline{x}(\lambda)$, $\overline{y}(\lambda)$, $\overline{z}(\lambda)$ are the color matching functions, $I(\lambda)$ is the spectral power distribution of the light source, $R(\lambda, \theta)$ are the reflected or transmitted light spectra of the object,
and $k$ is a normalizing factor defined as
\begin{equation}
k \equiv \int_{\lambda_1}^{\lambda_2} \overline{y}(\lambda) I(\lambda) \, d\lambda.
\label{eq_2.2}
\end{equation}

The CIE 1931 XYZ color space is preferred for chromaticity calculations because it avoids the occurrence of negative tristimulus values \cite{kerr2010_3_51, ohno2000_3_52, wyman2013_3_53}.
The CIE 1931 XYZ chromaticity diagram serves as a perceptual map to distinguish between different color stimuli. Each point within this diagram represents a specific color, and its coordinates $x$, $y$ are determined by
\begin{equation}
\left\{
\begin{aligned}
x &= \frac{X}{X + Y + Z}, \\
y &= \frac{Y}{X + Y + Z}.
\end{aligned}
\right.
\label{eq_2.3}
\end{equation}
For two points with coordinates ($x_1$, $y_1$) and ($x_2$, $y_2$), representing two different colors on the chromaticity diagram, the Euclidean distance $d_{1,2}$ between them quantifies the perceptual difference between the two colors:
\begin{equation}
d_{1,2} = \sqrt{(x_1 - x_2)^2 + (y_1 - y_2)^2}.
\label{eq_MF}
\end{equation}

In a multilayer structure with $l$ layers with thicknesses $\mathbf{t} = [t_1\, t_2 \, ... \, t_l]$ the material composition of the layers is represented by the vector $\mathbf{m} = [m_1 \, m_2 \, ...\, m_l]$, where $m_i \,(i\,=\,1,\,2,\,...\,,l)$ are integers between 1 and $M$, and $M$ is the number of materials considered. 
A multilayer structure containing a single phase-change material exhibits two distinct reflection spectra, depending on whether the PCM is in its crystalline (C) or amorphous (A) phase. Given a structure with a specific material composition $\mathbf{m}$ and layer thicknesses $\mathbf{t}$, the reflection spectra for the two phases of the phase-change material can be calculated using the impedance method \cite{haus1984waves}.
From these reflection spectra, we obtain two sets of tristimulus values for the structure, ($X_{C}$, $Y_{C}$, $Z_{C}$) and ($X_{A}$, $Y_{A}$, $Z_{A}$), using Eqs. (\ref{eq_2.1}) and (\ref{eq_2.2}) over the visible wavelength range. Using Eq. (\ref{eq_2.3}), we then obtain two sets of color coordinates for the multilayer structure, ($x_{C}$, $y_{C}$) and ($x_{A}$, $y_{A}$), on the CIE diagram, corresponding to the two phases of the phase-change material. 

Since our objective is to obtain two maximally distinct colors,
we maximize the merit function for two-color structures, $F_{\rm two-color}(\mathbf{m},\mathbf{t})$, defined as
\begin{equation}
F_{\rm two-color}(\mathbf{m},\mathbf{t}) =  d_{\rm C,A},
\label{eq_MF_2}
\end{equation}
where $d_{\rm C,A}$ is the distance between the two sets of color coordinates corresponding to the crystalline and amorphous phases of the phase-change material, calculated using Eq. (\ref{eq_MF}).

We use a memetic optimization algorithm to optimize both the material composition and the layer thicknesses of the aperiodic multilayer structures in order to maximize the merit function $F_{\rm two-color}$ [Eq. (\ref{eq_MF_2})], thereby generating two maximally distinct colors on the CIE diagram from a single multilayer structure which incorporates a phase-change material. If the optimized structure includes multiple layers of the phase-change material, we assume that all PCM layers are either in the crystalline or the amorphous phase to obtain the two distinct colors.

In the case of optimizing structures to generate four maximally distinct colors, we consider configurations with two GST layers to enable multistate phase combinations. To enforce this constraint, any structure in the evolutionary process that does not contain exactly two PCM layers is assigned a penalizing value for its fitness function. This ensures that such structures are excluded from selection as the final optimized design.

A four-color structure supports four distinct states, each corresponding to one of the possible phase combinations of the two GST layers: both layers crystalline (CC), both layers amorphous (AA), top layer crystalline and bottom layer amorphous (CA), and top layer amorphous with bottom layer crystalline (AC). Each of these states maps to a unique point on the chromaticity diagram. These four points define six pairwise distances: $d_{\rm CC,AA}$, $d_{\rm CC,CA}$, $d_{\rm CC,AC}$, $d_{\rm AA,CA}$, $d_{\rm AA,AC}$, and $d_{\rm CA,AC}$. As before, these distances are calculated using the Euclidean formula [Eq. (\ref{eq_MF})]. To achieve maximally distinct colors, we maximize the merit function for four-color structures, $F_{\rm four-color}(\mathbf{m},\mathbf{t})$, defined as
\begin{equation}
F_{\rm four-color}(\mathbf{m},\mathbf{t}) = \min (d_{\rm CC,AA}, d_{\rm CC,CA}, d_{\rm CC,AC}, d_{\rm AA,CA}, d_{\rm AA,AC}, d_{\rm CA,AC}).
\label{eq_MF_4}
\end{equation}
In other words, we aim to maximize the minimum pairwise distance among all six possible distances between the four color points on the CIE chromaticity diagram.

Here, the material of each layer is selected from a set of 18 candidates ($M=18$), which include dielectrics: aluminum oxide (Al$_2$O$_3$), hafnium oxide (HfO$_2$), magnesium fluoride (MgF$_2$), silicon carbide (SiC), silicon nitride (Si$_3$N$_4$), silicon dioxide (SiO$_2$), tantalum pentoxide (Ta$_2$O$_5$), titanium dioxide (TiO$_2$), indium tin oxide (ITO), aluminum nitride (AlN), and titanium nitride (TiN); metals: aluminum (Al), silver (Ag), tungsten (W), and chromium (Cr); semiconductors: silicon (Si) and germanium (Ge); a phase-change material. We use silver (Ag) as the substrate, while the superstrate is air. To enable color tunability, we consider several PCMs with distinct optical properties in their two phases. These include germanium-antimony-tellurium (GST), vanadium dioxide (VO$_2$), germanium-antimony-selenium-tellurium (GSST), silver-indium-antimony-tellurium (AIST), antimony trisulfide (Sb$_2$S$_3$), germanium telluride (GeTe), and molybdenum oxide (MoO$_x$). 
Among these, GST, a widely used phase-change material in photonics applications, demonstrated the best performance in terms of color distinctiveness. Hence, in the remainder of this paper, we focus on structures utilizing GST as the PCM. 
For all materials, we use experimentally measured wavelength-dependent refractive index data \cite{palik1998_5-29, kischkat2012_5-30, nature_date, GST_data_MIT:18}.

In this paper, we employ a memetic optimization algorithm to optimize structural colors \cite{fan2017}. Initially, we generate a random population of $N$ $l$-layer structures. During crossover, we pair individuals to produce offspring by combining sections of the multilayer designs of the parents. A mutation step then randomly alters the material or thickness of a layer within a randomly selected structure. The next generation is selected based on the merit function, ensuring diversity by choosing a certain percentage of individuals from both the top and bottom of the population. Elite structures undergo refinement through local optimization of layer thicknesses using the quasi-Newton method. 
To enhance the ability of the algorithm to escape local minima and converge to the global optimum, we introduce a modification beyond the approach described in Ref. \cite{fan2017}. Upon convergence, the best individual from the final population is retained, while the rest of the population is re-initialized with randomly generated structures. Details of this modification, as well as the parameters used in the algorithm, can be found in our previous work \cite{emrose2024}.

\section{Results}
\label{Results_2}

\begin{figure}[htb!]
\centering\includegraphics[width=11cm]{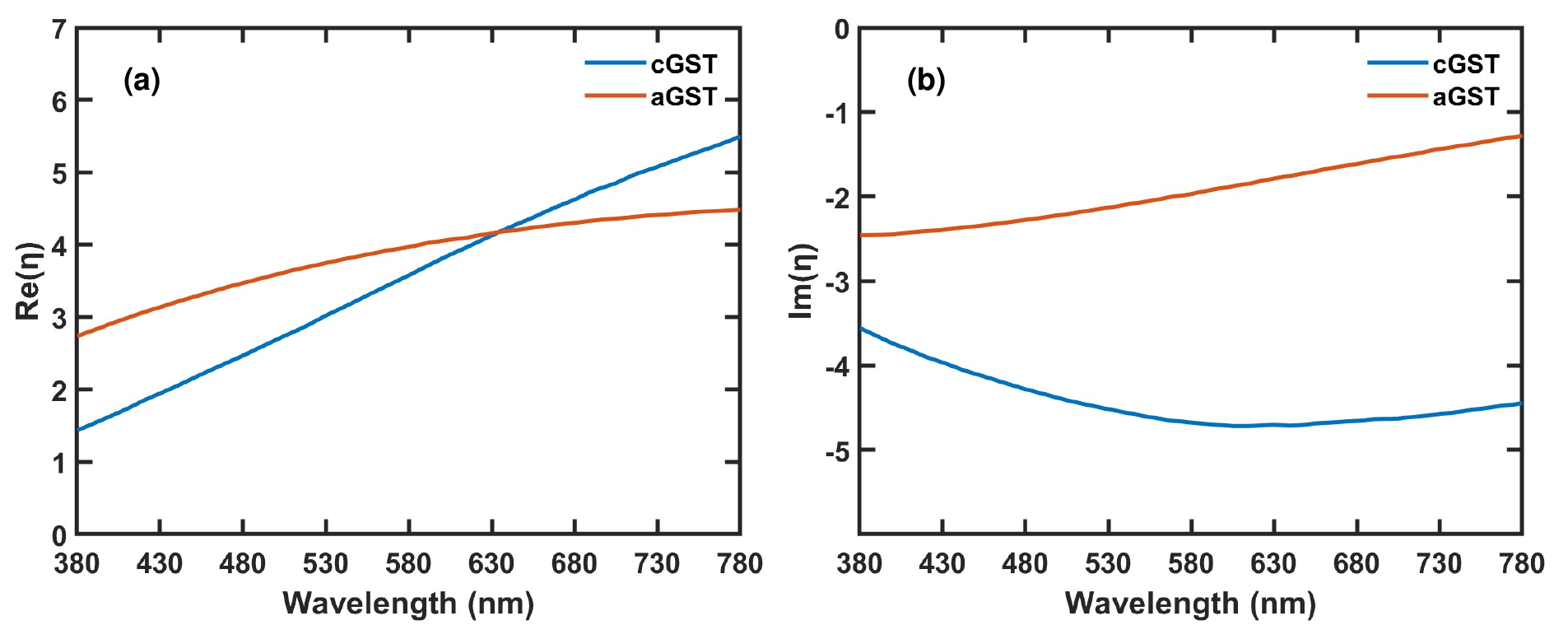} 
\caption{(a) Real and (b) imaginary part of the refractive index of GST in its amorphous (aGST) and crystalline (cGST) phases in the visible wavelength range.}
\label{Fig_Refractive_Indecies}
\end{figure}

The optical properties of Ge$_2$Sb$_2$Te$_5$ (GST) in its amorphous (aGST) and crystalline (cGST) phases are significantly different. External electrical pulses, optical pulses, or thermal annealing can rapidly and reversibly induce the transition between the two phases, with power consumption occurring only during the transition. Figs. \ref{Fig_Refractive_Indecies}(a) and \ref{Fig_Refractive_Indecies}(b) show the real and imaginary parts, respectively, of the refractive index of GST in both its amorphous and crystalline phases within the visible wavelength range \cite{GST_data_MIT:18}. Fig. \ref{Fig_Refractive_Indecies}(a) shows that the real parts of the refractive indices of cGST and aGST differ markedly. Furthermore, cGST exhibits greater optical loss than aGST, as demonstrated in Fig. \ref{Fig_Refractive_Indecies}(b). These differences in refractive indices enable the generation of distinct structural colors when GST is switched between its two phases.

Here, we introduce multilayer structures based on GST for reconfigurable structural color generation. For both two-color and four-color designs, our goal is to generate the most visually distinct colors on the CIE chromaticity diagram. We optimize the structures for normally incident light ($\theta=0^\circ$). Additionally, to eliminate structures with low reflection, we impose a constraint in our optimization algorithm that the resulting structure must exhibit a maximum reflection of at least 40\% within the wavelength range of interest for both GST phases. 

\subsection{Two-color structures}
\label{2color}

We first employ the memetic optimization algorithm described in Section \ref{Theory_2} to design a multilayer structure with at least one GST layer that can produce two distinct colors by switching between the two phases of GST. We optimize both the material composition and the layer thicknesses of the structures to maximize the distance between the coordinates of the two colors on the CIE chromaticity diagram [Eq. (\ref{eq_MF})] for normally incident light.

\begin{figure}[htb!]
\centering
\centering\includegraphics[width=11cm]{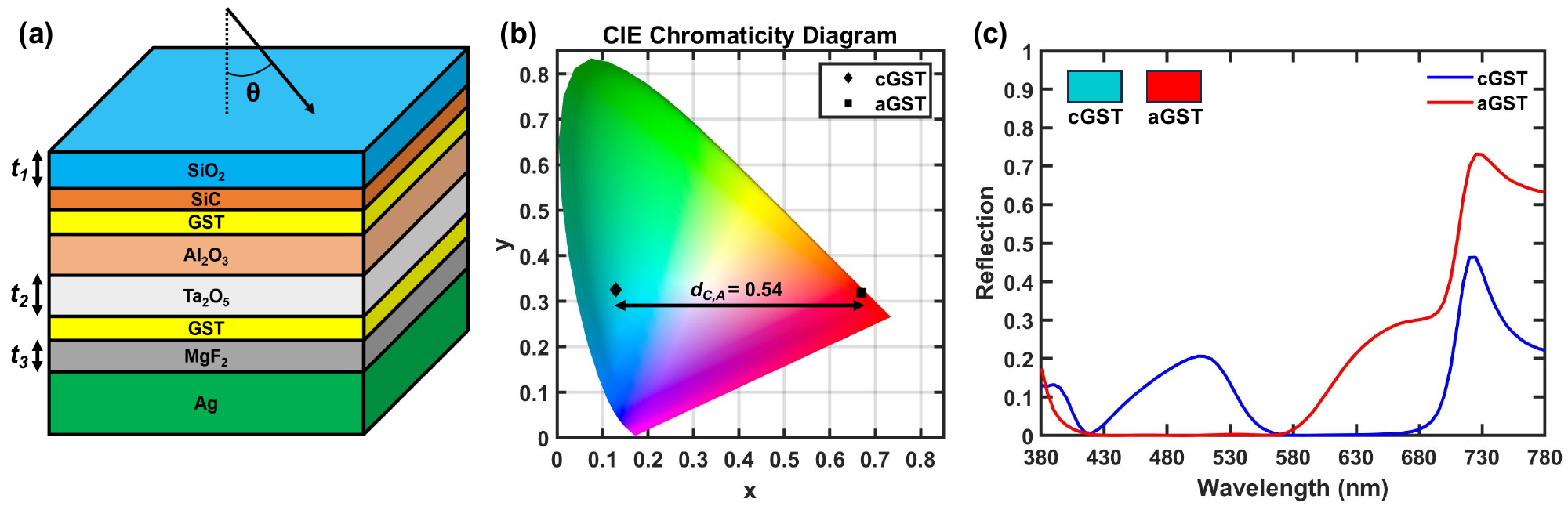} 
\caption{(a) Schematic showing the optimized material composition of seven-layer structures above a silver substrate  obtained from the memetic optimization algorithm for two maximally distinct colors. The thicknesses of the SiO$_2$, Ta$_2$O$_5$, and MgF$_2$ layers are denoted as \textit{t$_1$}, \textit{t$_2$}, and \textit{t$_3$}, respectively. (b) CIE chromaticity diagram showing the two colors produced by the optimized structure when the two GST layers are in either the crystalline (cGST) or amorphous (aGST) phase. Here, we maximize the Euclidean distance between the coordinates of the two colors on the CIE diagram, achieving a separation of 0.54. (c) Reflection as a function of wavelength for the optimized seven-layer structure with GST in its crystalline and amorphous phases. Results are shown for normally incident light. The material and thickness of each layer in the optimized structure are given in Table \ref{table:2.one}. The corresponding color for each phase is shown in the upper left corner of the figure.}
\label{Fig_2color_Structure_CIE_Reflection}
\end{figure}

Fig. \ref{Fig_2color_Structure_CIE_Reflection}(a) shows the optimized material composition of seven-layer structures above a silver substrate obtained from the memetic optimization algorithm for two maximally distinct colors. Table \ref{table:2.one} lists the material and thickness of each layer in the optimized structure depicted in Fig. \ref{Fig_2color_Structure_CIE_Reflection}(a). The design consists of two thin GST films separated by two dielectric spacer layers, Al$_2$O$_3$ and Ta$_2$O$_5$ (Table \ref{table:2.one}). It also includes two additional dielectric layers (SiO$_2$ and SiC) on top of the first GST layer, as well as a MgF$_2$ layer between the second GST layer and the silver substrate. To obtain two distinct colors, we consider two different states of the structure. In the first state both GST layers are in the crystalline phase, while in the second state they are in the amorphous phase.
In addition to its color tunability, the proposed multilayer structure eliminates the need for subwavelength lithography, commonly required in metasurface-based color filters, making it more suitable for large-scale applications \cite{jafari_2019_1}.

Fig. \ref{Fig_2color_Structure_CIE_Reflection}(b) shows the corresponding colors of the structure in the CIE diagram when the two GST layers are in either the crystalline or amorphous phase. In the optimization procedure, we maximize the distance between these two points. The maximum distance obtained in the CIE diagram is $\sim$0.54.
Since both points are located near the edge of the chromaticity diagram, the resulting color separation is large.
From the CIE diagram, we observe that the structure appears red when both GST layers are in the amorphous phase and turquoise when both are in the crystalline phase. We also show the reflection spectra of the structure corresponding to each color in Fig. \ref{Fig_2color_Structure_CIE_Reflection}(c). Both phases exhibit a resonance at $\sim$730 nm. However, an additional resonance appears at $\sim$510 nm for the crystalline phase, which distinguishes the two colors. The corresponding color for each phase is shown in the upper left corner of Fig. \ref{Fig_2color_Structure_CIE_Reflection}(c).

\begin{table}[ht]
    \centering
    \caption{Material and Thickness of Each Layer in the Optimized Structure of Fig. \ref{Fig_2color_Structure_CIE_Reflection}(a)}
    \begin{tabular}{c|c|c|}
        \toprule
        Layer   & Material       & Thickness (nm)\\
        \midrule
                & Air            & Superstrate\\
        1       & SiO$_2$        & 74\\
        2       & SiC            & 39\\
        3       & GST            & 9\\
        4       & Al$_2$O$_3$    & 366\\
        5       & Ta$_2$O$_5$    & 69\\
        6       & GST            & 13\\
        7       & MgF$_2$        & 224\\
                & Ag             & Substrate\\
        \bottomrule
    \end{tabular}
    \label{table:2.one}
\end{table}

\begin{figure}[htb!]
\centering
\centering\includegraphics[width=11cm]{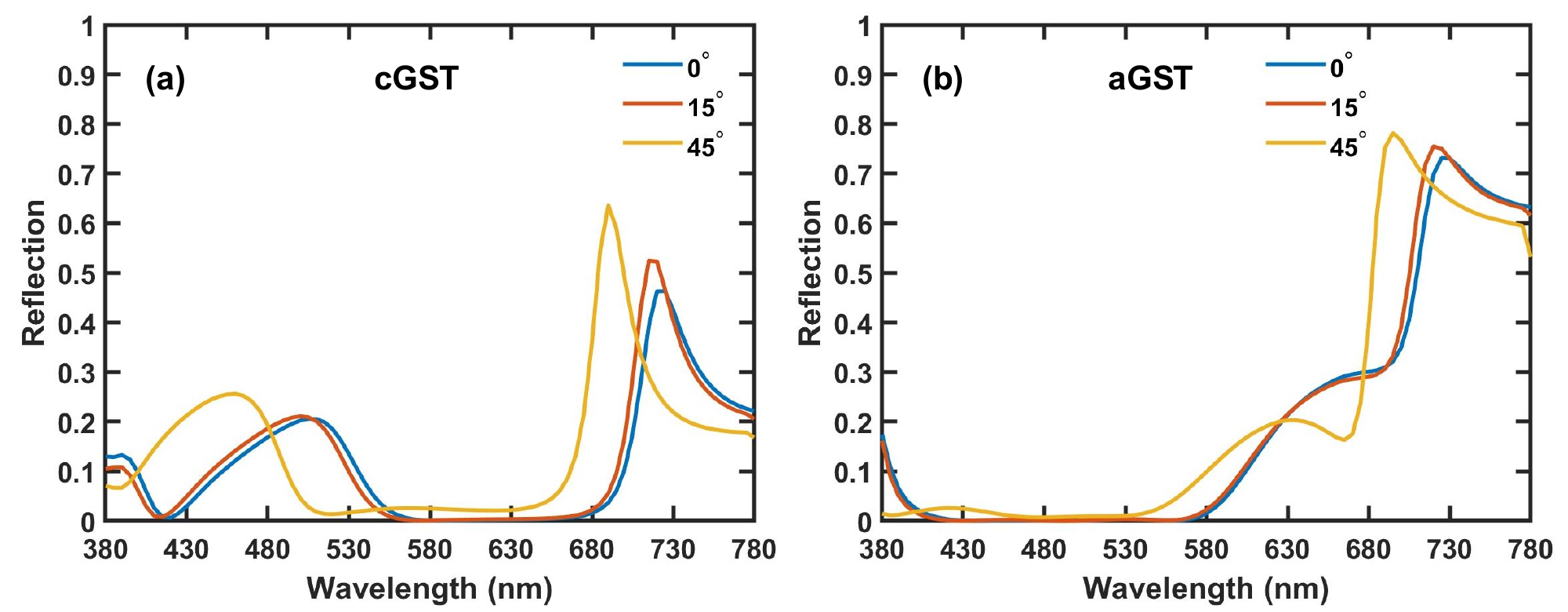} 
\caption{Reflection spectra of the structure shown in Fig. \ref{Fig_2color_Structure_CIE_Reflection}(a) in the visible wavelength range at incident angles of $0^\circ$, $15^\circ$, and $45^\circ$ with both GST layers in the (a) crystalline (b) amorphous phase. All other parameters are as given in Table~\ref{table:2.one}.}
\label{Fig_2color_Angular_Response}
\end{figure}

We also calculate the angular response of the optimized structure of Table~\ref{table:2.one}. In Fig. \ref{Fig_2color_Angular_Response}(a) [Fig. \ref{Fig_2color_Angular_Response}(b)], we show the reflection spectra for different incident angles when both GST layers in the optimized structure are in the crystalline (amorphous) phase. In both cases, we observe that, as the incident angle increases, the reflection resonances shift to shorter wavelengths. This shift is very small for incident angles up to $15^\circ$, 
but becomes significantly more pronounced beyond this angle.

\begin{figure}[htb!]
\centering
\centering\includegraphics[width=11cm]{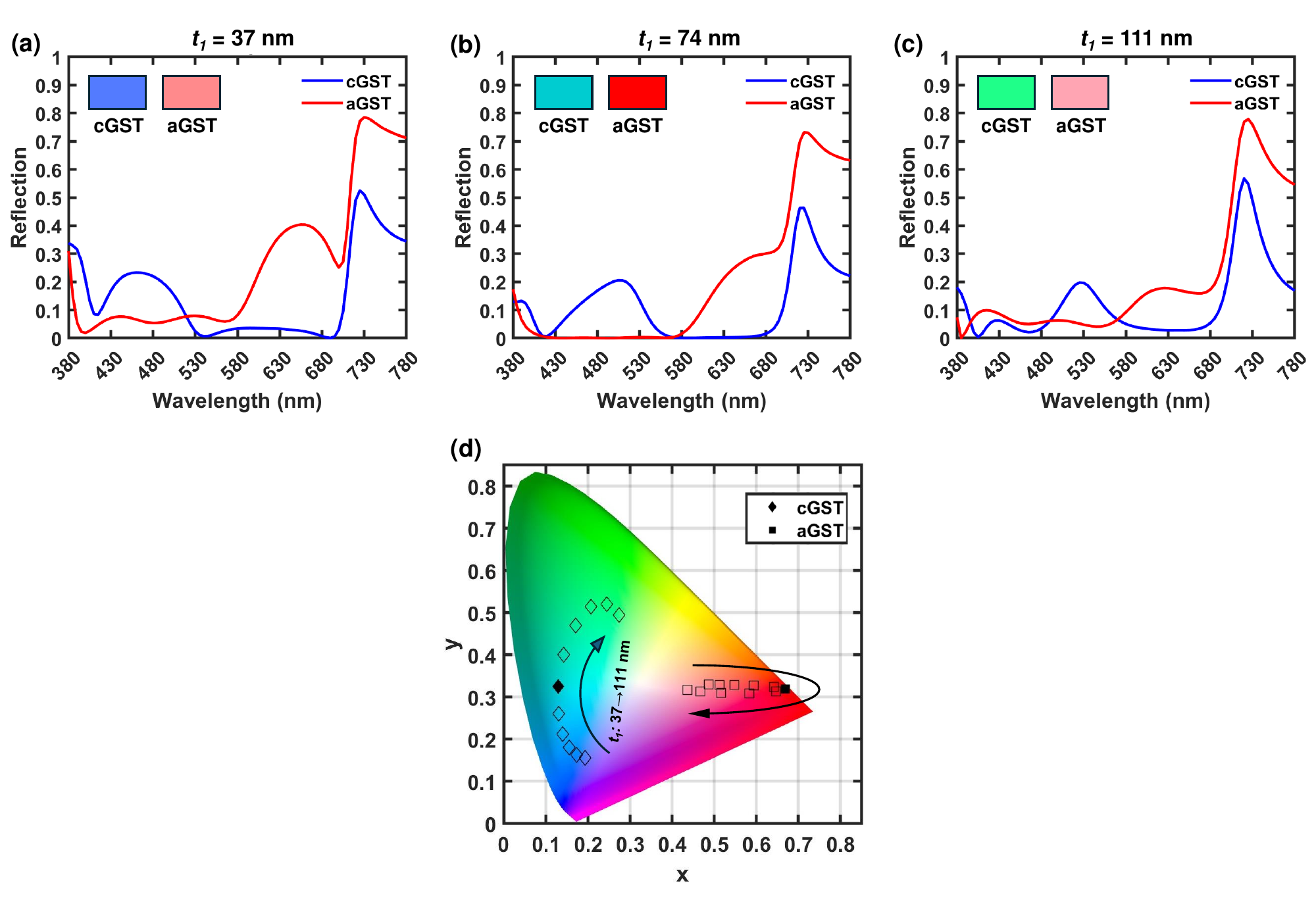} 
\caption{Reflection as a function of wavelength in the visible range when the thickness of the SiO$_2$ layer, \textit{t$_1$}, is (a) 37 nm, (b) 74 nm,  and (c) 111 nm. 
All other layer thicknesses are kept at their optimized values, as listed in Table ~\ref{table:2.one}. Results are shown for normally incident light.
Each subfigure includes the reflection spectra corresponding to the crystalline (cGST) and amorphous (aGST) phases of the two GST layers. 
The associated structural colors produced by the structure for each phase are shown in the top-left corner of each subfigure.
(d) CIE chromaticity diagram showing the colors produced by the structure as the thickness of the SiO$_2$ layer,  \textit{t$_1$},  varies from 37 nm to 111 nm. The arrows indicate the direction of the structural colors shift as \textit{t$_1$} increases. The filled symbols for cGST and aGST correspond to the optimized value for two maximally distinct colors of \textit{t$_1$}=74 nm.}
\label{Fig_2color_Thickness_Response_SiO2}
\end{figure}

To elucidate the role of a specific layer in the optimized structure, we examine the reflection spectra corresponding to the two phases of GST as the thickness of that layer is varied.
Figs. \ref{Fig_2color_Thickness_Response_SiO2}(a), \ref{Fig_2color_Thickness_Response_SiO2}(b), and \ref{Fig_2color_Thickness_Response_SiO2}(c) illustrate how the reflection spectra of the structure change as the thickness of the SiO$_2$ layer adjacent to air, \textit{t$_1$} [Fig. \ref{Fig_2color_Structure_CIE_Reflection}(a)], is varied, while all other layer thicknesses are maintained at their optimized values (Table 1).
When \textit{t$_1$} varies from 37 nm to 111 nm, the short-wavelength resonance peak in the crystalline phase of GST shifts from $\sim$460 nm to $\sim$530 nm. Consequently, the color produced by the structure in the crystalline phase varies from blue to light green [Fig. \ref{Fig_2color_Thickness_Response_SiO2}(d)].
In addition, as \textit{t$_1$} varies from 37 nm to 111 nm, the greatest contrast in the reflection spectra between the two GST phases within the 420-680 nm wavelength range is achieved for \textit{t$_1$}=74 nm [Fig. \ref{Fig_2color_Thickness_Response_SiO2}(b)]. As a result, the two most distinct colors corresponding to the amorphous and crystalline phases of GST are achieved when \textit{t$_1$}=74 nm [Fig. \ref{Fig_2color_Thickness_Response_SiO2}(d)].


\begin{figure}[htb!]
\centering
\centering\includegraphics[width=11cm]{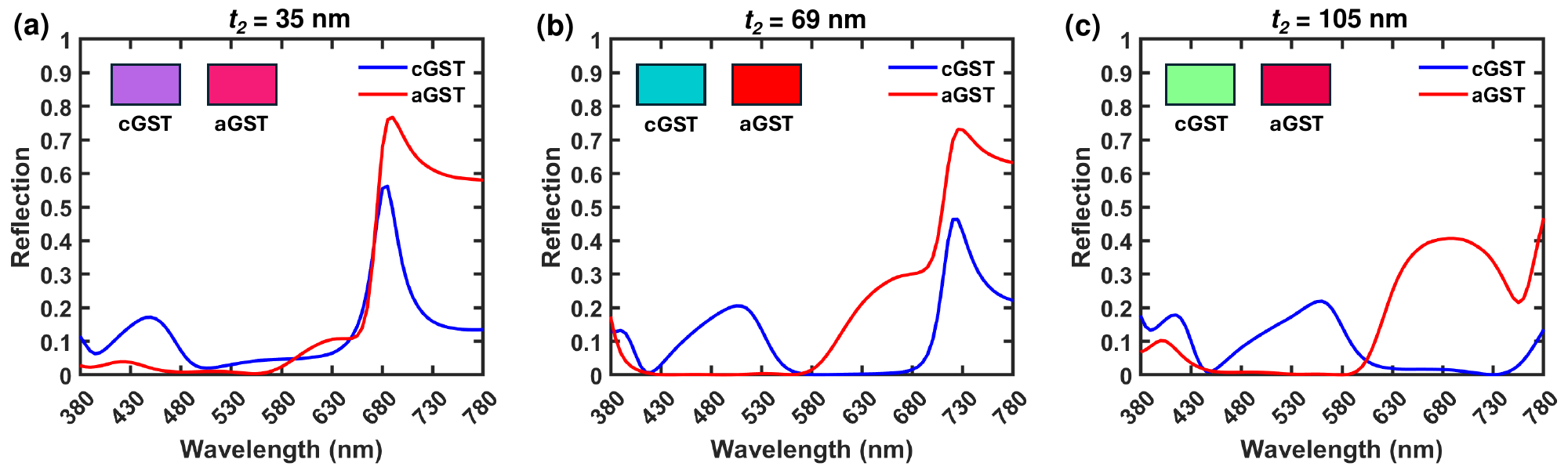} 
\caption{Reflection as a function wavelength in the visible range when the thickness of the Ta$_2$O$_5$ layer, \textit{t$_2$}, is (a) 35 nm, (b) 69 nm,  and (c) 105 nm. All other layer thicknesses are kept at their optimized values, as listed in Table ~\ref{table:2.one}. Results are shown for normally incident light. 
Each subfigure includes the reflection spectra corresponding to the
crystalline (cGST) and amorphous (aGST) phases of the two GST layers. The associated structural
colors produced by the structure for each phase are shown in the top-left corner of each
subfigure.}
\label{Fig_2color_Thickness_Response_Ta2O5}
\end{figure}

We also analyze the reflection spectra corresponding to the two phases of GST under varying thicknesses of the Ta$_2$O$_5$ dielectric spacer layer between the two GST layers, denoted as \textit{t$_2$} [Fig. \ref{Fig_2color_Structure_CIE_Reflection}(a)]. Figures \ref{Fig_2color_Thickness_Response_Ta2O5}(a), \ref{Fig_2color_Thickness_Response_Ta2O5}(b), and \ref{Fig_2color_Thickness_Response_Ta2O5}(c) show the spectra for \textit{t$_2$} = 35 nm, 69 nm, and 105 nm, respectively. As \textit{t$_2$} increases from 35 nm to 105 nm, the short-wavelength resonance peak in the crystalline phase (cGST) shifts from $\sim$450 nm to $\sim$560 nm, tuning the structural color from purple toward light green. The dominant peak in the amorphous phase (aGST), located at $\sim$780 nm, exhibits negligible wavelength shift between \textit{t$_2$} = 35 nm [Fig. \ref{Fig_2color_Thickness_Response_Ta2O5}(a)] and \textit{t$_2$} = 69 nm [Fig. \ref{Fig_2color_Thickness_Response_Ta2O5}(b)], but its amplitude decreases for \textit{t$_2$} = 105 nm [Fig. \ref{Fig_2color_Thickness_Response_Ta2O5}(c)]. Since amplitude variations do not affect the perceived color \cite{wang2023_3}, the aGST color remains essentially unchanged across this thickness range. At \textit{t$_2$} = 69 nm [Fig. \ref{Fig_2color_Thickness_Response_Ta2O5}(b)], the spectral separation in the 420–680 nm range, and thus the color contrast, between cGST and aGST reaches its maximum.


\begin{figure}[htb!]
\centering
\centering\includegraphics[width=11cm]{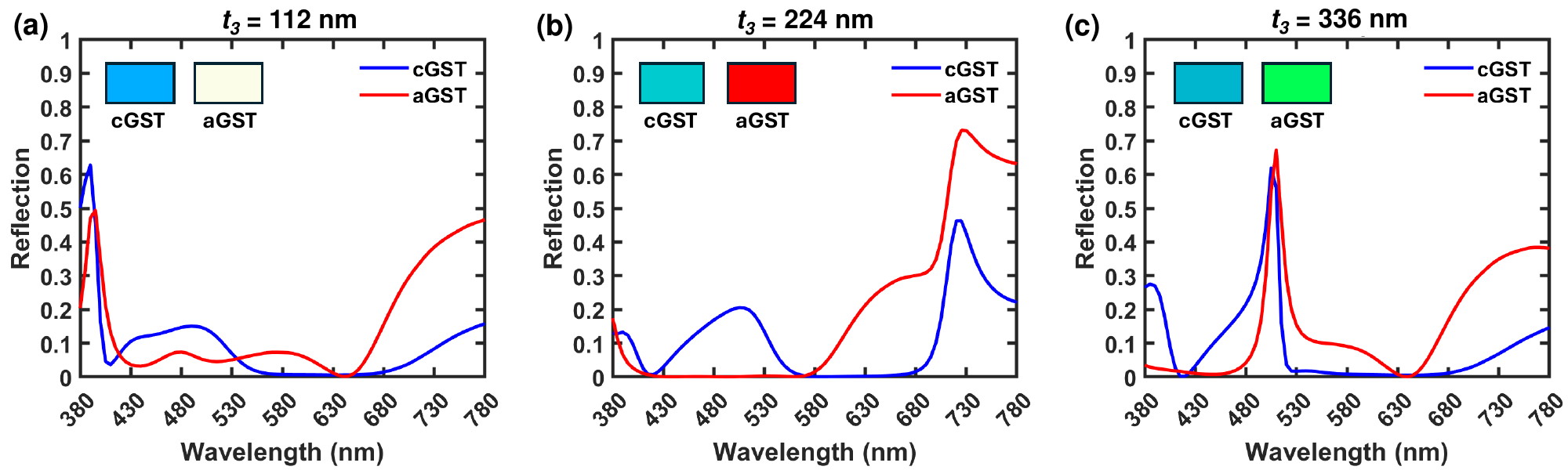} 
\caption{Reflection as a function wavelength in the visible range when the thickness of the MgF$_2$ layer, \textit{t$_3$}, is (a) 112 nm, (b) 224 nm,  and (c) 336 nm. All other layer thicknesses are kept at their optimized values, as listed in Table ~\ref{table:2.one}. Results are shown for normally incident light. 
Each subfigure includes the reflection spectra corresponding to the
crystalline (cGST) and amorphous (aGST) phases of the two GST layers. The associated structural
colors produced by the structure for each phase are shown in the top-left corner of each
subfigure.}
\label{Fig_2color_Thickness_Response_MgF2}
\end{figure}

We also investigate the influence of the thickness of the MgF$_2$ spacer layer between the second GST layer and the silver substrate, denoted as \textit{t$_3$} [Fig. \ref{Fig_2color_Structure_CIE_Reflection}(a)]. Figures \ref{Fig_2color_Thickness_Response_MgF2}(a), \ref{Fig_2color_Thickness_Response_MgF2}(b), and \ref{Fig_2color_Thickness_Response_MgF2}(c) show the reflection spectra for both crystalline (cGST) and amorphous (aGST) phases for \textit{t$_3$} = 112 nm, 224 nm, and 336 nm, respectively, while all other layer thicknesses are kept at their optimized values (Table 1). In this scenario, the short-wavelength resonance peak of the cGST phase at $\sim$510 nm does not shift significantly as \textit{t$_3$} increases from 112 nm to 336 nm. However, its linewidth narrows and the peak reflectance increases, reaching $\sim$70\% at \textit{t$_3$} = 336 nm [Fig. \ref{Fig_2color_Thickness_Response_MgF2}(c)]. A similar narrowing and increase in amplitude are observed for the aGST resonance at the same wavelength. The greatest contrast in the reflection spectra between the crystalline and amorphous phases occurs at the optimized value of \textit{t$_3$} = 224 nm [Fig. \ref{Fig_2color_Thickness_Response_MgF2}(b)].


\subsection{Four-color structures}
\label{4color}

In addition to designing structures which generate two colors, we apply the memetic optimization algorithm described in Section \ref{Theory_2} to optimize multilayer structures for producing four maximally distinct colors. We optimize both the material composition and the layer thicknesses of the structures to maximize the minimum pairwise distance between the four colors on the CIE chromaticity diagram [Eq. (\ref{eq_MF_4})].

\begin{figure}[htb!]
\centering
\centering\includegraphics[width=11cm]{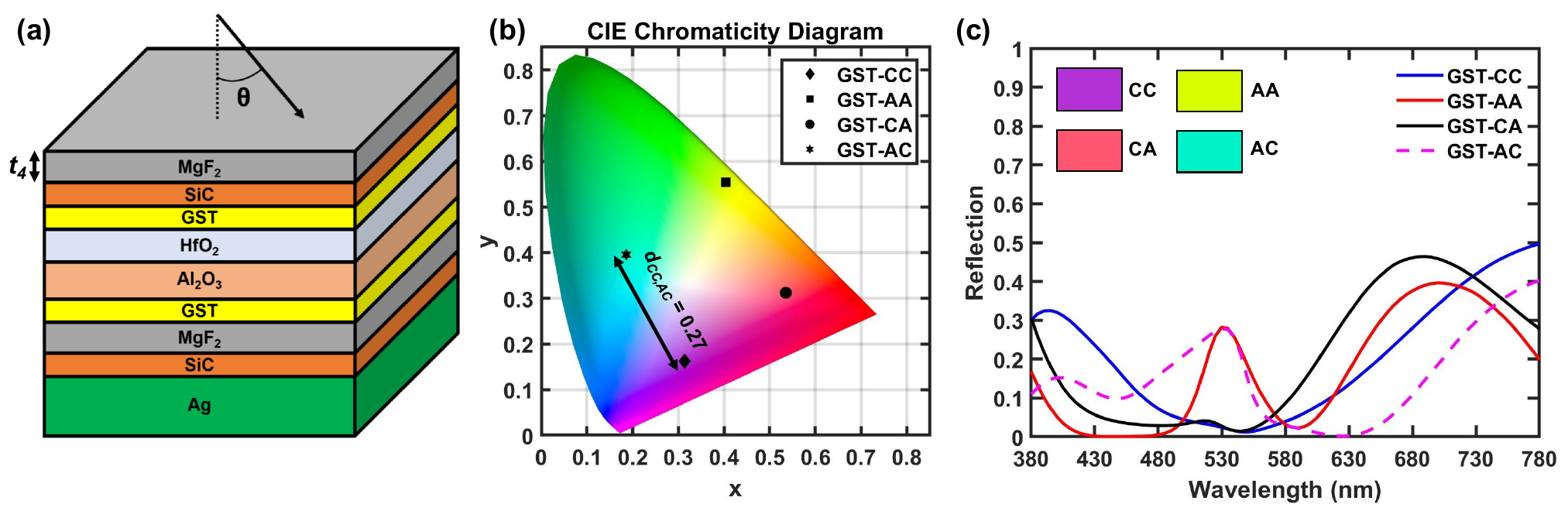} 
\caption{(a) Schematic showing the optimized material composition of an eight-layer structure above a silver substrate for four maximally distinct colors. The thickness of the top MgF$_2$ layer is denoted as \textit{t$_4$}. (b) CIE chromaticity diagram showing the four colors produced by the optimized structure corresponding to the four different combinations of the phases of the two GST layers: both layers crystalline (CC), both layers amorphous (AA), top layer crystalline and bottom layer amorphous (CA), and top layer amorphous with bottom layer crystalline (AC). Here, we maximize the minimum Euclidean distance between the coordinates of the four colors on the CIE diagram. The maximized minimum distance achieved in the optimized structure is 0.27 units, occuring between the CC and AC states. (c) Reflection as a function of wavelength for the optimized eight-layer structure. Shown are the reflection spectra corresponding to the four different combinations of the phases of the two GST layers (CC, AA, CA, AC). The associated structural colors produced by the structure for each of the four combinations are shown in the top-left corner. Results are shown for normally incident light. The material and thickness of each layer in the optimized structure are given in Table \ref{table:2.two}.}
\label{Fig_4color_Structure_CIE_Reflection}
\end{figure}

Figure \ref{Fig_4color_Structure_CIE_Reflection}(a) shows the optimized material composition for an eight-layer structure above a silver substrate, as determined by the memetic optimization algorithm to yield four maximally distinct colors. 
Table~\ref{table:2.two} provides the material and thickness of each layer in the optimized structure shown in Fig. \ref{Fig_4color_Structure_CIE_Reflection}(a). The design features two thin GST layers: the top GST layer is situated beneath layers of MgF$_2$ and SiC. The bottom GST layer is placed above layers of MgF$_2$ and SiC on top of the silver substrate. Additionally,  dielectric layers of HfO$_2$ and Al$_2$O$_3$ separate the two GST layers.

Figure \ref{Fig_4color_Structure_CIE_Reflection}(b) shows the four colors produced by the structure on the CIE chromaticity diagram, corresponding to the four possible phase combinations of the two GST layers. The memetic optimization algorithm aims to maximize the distinction among these colors. Figure \ref{Fig_4color_Structure_CIE_Reflection}(c) shows the reflection spectra of the optimized structure in its four phase states. The corresponding structural colors for each phase combination are shown in the top-left corner of the figure. The CC state (blue curve) exhibits two peaks at $\sim$400 nm and $\sim$780 nm, corresponding to a blue color. The AA state (red curve) displays three peaks at $\sim$380 nm, $\sim$530 nm, and $\sim$700 nm, resulting in a yellow color. The CA state (black curve) produces a red color with resonant peaks at $\sim$380 nm and $\sim$680 nm. Finally, the AC state (dashed line) yields a cyan color with peaks at $\sim$400 nm, $\sim$530 nm, and $\sim$780 nm. The minimum distance between any two color coordinates is $\sim$0.27, indicating that the optimized structure successfully generates four highly distinct colors. This color contrast performance exceeds that of previously reported structures in the literature \cite{jafari_2019_1}.

\begin{table}[ht]
    \centering
    \caption{Material and Thickness of Each Layer in the Optimized Structure of Fig. ~\ref{Fig_4color_Structure_CIE_Reflection}(a)}
    \begin{tabular}{c|c|c|}
        \toprule
        Layer   & Material       & Thickness (nm)\\
        \midrule
                & Air            & Superstrate\\
        1       & MgF$_2$        & 86\\
        2       & SiC            & 127\\
        3       & GST            & 6\\
        4       & HfO$_2$        & 90\\
        5       & Al$_2$O$_3$    & 137\\
        6       & GST            & 10\\
        7       & MgF$_2$        & 76\\
        8       & SiC            & 139\\
                & Ag             & Substrate\\
        \bottomrule
    \end{tabular}
    \label{table:2.two}
\end{table}

\begin{figure}[htb!]
\centering
\centering\includegraphics[width=11cm]{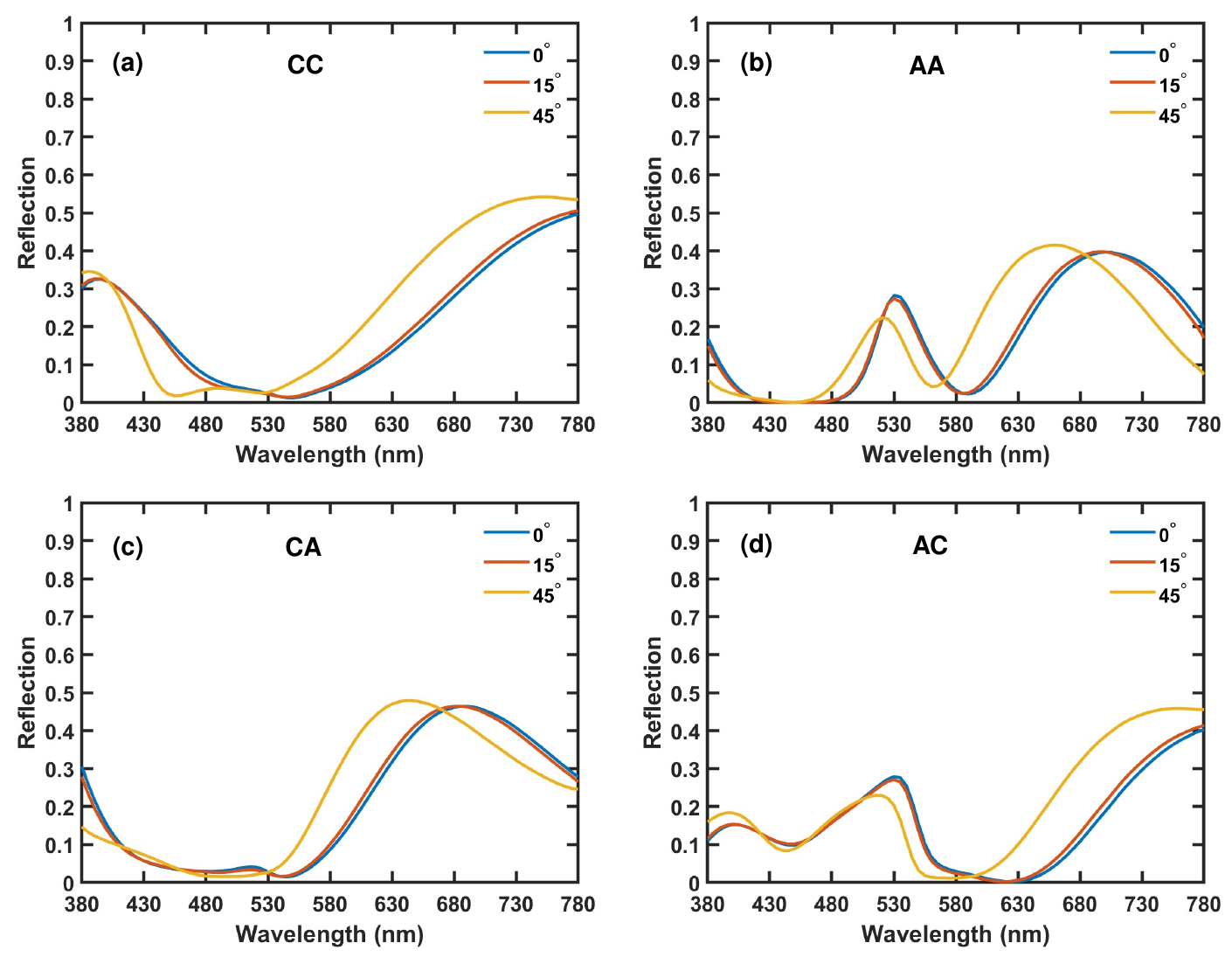} 
\caption{Reflection spectra of the structure shown in Fig. \ref{Fig_4color_Structure_CIE_Reflection}(a) in the visible wavelength range at incident angles of $0^\circ$, $15^\circ$, and $45^\circ$, when (a) both GST layers are in the crystalline phase (CC), (b) both layers are in the amorphous phase (AA), (c) the top layer is in the crystalline and the bottom layer is in the amorphous phase (CA), and (d) the top layer is in the amorphous phase with the bottom layer in the crystalline phase (AC). All other parameters are as given in Table~\ref{table:2.two}.}
\label{Fig_4color_Angular_Response}
\end{figure}

We also investigate how the reflection spectra of our optimized structure are sensitive to the incident angle. In Fig. \ref{Fig_4color_Angular_Response}, we show the reflection spectra as the incident angle increases from $0^\circ$ to $15^\circ$ and $45^\circ$, for all four phase states (CC, AA, CA, and AC). Similar to the two-color structures (Fig. \ref{Fig_2color_Angular_Response}), the reflection spectra in all four states remain nearly identical to those at normal incidence for incident angles up to $15^\circ$. However, beyond this angle, we observe a significant shift in the reflection spectra. In all cases, the resonance peaks shift to shorter wavelengths as the incident angle increases. 
For instance, in the CC case [Fig. \ref{Fig_4color_Angular_Response}(a)] the resonance peak at $\sim$780 nm shifts to $\sim$730 nm as the incident angle increases from $0^\circ$ to $45^\circ$. 
The other three states (AA, CA, and AC), exhibit similar angular dependence in their reflection spectra.

\begin{figure}[htb!]
\centering
\centering\includegraphics[width=11cm]{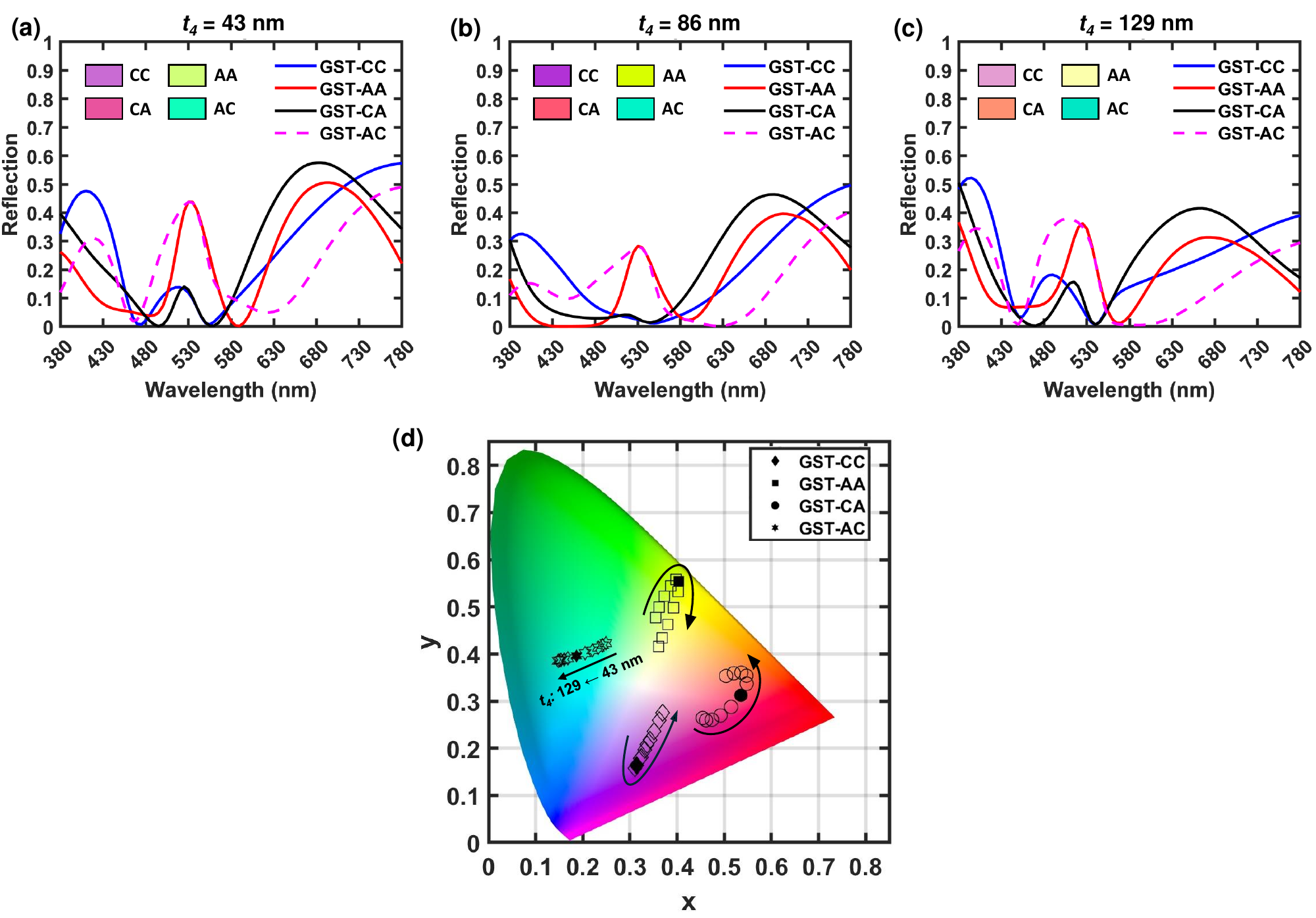} 
\caption{Reflection as a function of wavelength in the visible range when the thickness of the top MgF$_2$ layer, \textit{t$_4$}, is (a) 43 nm, (b) 86 nm,  and (c) 129 nm. 
All other layer thicknesses are kept at their optimized values, as listed in Table ~\ref{table:2.two}. Results are shown for normally incident light.
Each subfigure includes the reflection spectra corresponding to the four different combinations of the phases of the two GST layers (CC, AA, CA, AC). The associated structural colors produced by the structure for each of the four combinations are shown in the top-left corner of each subfigure. (d) CIE chromaticity diagram showing the colors produced by the structure as the thickness of the top MgF$_2$ layer,  \textit{t$_4$}, varies from 43 nm to 129 nm. The arrows indicate the direction of the structural colors shift as \textit{t$_4$} increases. The filled symbols for CC, AA, CA, and AC correspond to the optimized value of \textit{t$_4$}=86 nm for four maximally distinct colors.}
\label{Fig_4color_Thickness Response}
\end{figure}

To elucidate the role of a specific layer in the optimized structure, we examine the reflection spectra corresponding to the four phase states (CC, AA, CA, and AC) as the thickness of that layer is varied.
In Fig. \ref{Fig_4color_Thickness Response}, we vary only the thickness of the top MgF$_2$ layer, \textit{t$_4$} [Fig. \ref{Fig_4color_Structure_CIE_Reflection}(a)], while keeping all other layers at their optimized values from Table \ref{table:2.two}. This allows us to investigate how changes in \textit{t$_4$} affect the reflection spectra and color contrast. As the thickness of the top MgF$_2$ layer, \textit{t$_4$}, increases from 43 nm to 129 nm, the resonance peaks of each state shift in different directions [Figs. \ref{Fig_4color_Thickness Response}(a), \ref{Fig_4color_Thickness Response}(b), and \ref{Fig_4color_Thickness Response}(c)]. These spectral shifts determine the trajectories traced on the CIE chromaticity diagram in Fig. \ref{Fig_4color_Thickness Response}(d), showing how the perceived colors evolve with increasing \textit{t$_4$}. 
We observe that for \textit{t$_4$}=86 nm, which is the optimized thickness for achieving four maximally distinct colors (Table \ref{table:2.two}), the colors corresponding to the CC and AA phase states lie close to the edge of the chromaticity diagram [Fig. \ref{Fig_4color_Thickness Response}(d)]. Thus, \textit{t$_4$}=86 nm leads to four maximally distinct colors.

\section{Conclusions}
\label{Conclusion_2}

In this paper, we introduced multilayer structures based on phase-change materials for reconfigurable structural color generation. For both two-color and four-color designs, our objective was to maximize the perceived color difference between the generated colors, thereby enhancing their visual distinctiveness to the human eye.
We employed a memetic optimization algorithm to simultaneously optimize the material composition and layer thicknesses of multilayer structures incorporating phase-change materials in order to maximize the merit function of Eq.~(\ref{eq_MF_2}) for two-color, and Eq.~(\ref{eq_MF_4}) for four-color structures with maximally distinct colors.

In the case of optimizing structures to generate two maximally distinct colors, we considered multilayer structures containing at least one phase-change material layer that can produce two distinct colors by switching between the two phases of the material. We maximized the distance between the coordinates of the two colors on the CIE chromaticity diagram. In the case of optimizing structures to generate four maximally distinct colors,
we considered configurations with exactly two PCM layers to enable multistate phase combinations,
and maximized the minimum pairwise distance among all six possible color pairs on the CIE diagram. We optimized all structures for normally incident light. 
The material of each layer was selected from a pool of 18 candidates, including dielectrics, metals, semiconductors, and a PCM. We used silver as the substrate, while the superstrate was air. We considered several PCMs and, among these, GST demonstrated the best performance in terms of color distinctiveness. Thus, we focused on structures utilizing GST as the PCM. 

For two maximally distinct colors, our optimized seven-layer design comprises two thin GST films separated by dielectric spacer layers of Al$_2$O$_3$ and Ta$_2$O$_5$. It also includes two additional dielectric layers (SiO$_2$ and SiC) atop the first GST layer, and a MgF$_2$ layer between the second GST layer and the silver substrate. 
The distance between the coordinates of the two generated colors in the CIE diagram is $\sim$0.54 with both color points located near the edge of the chromaticity diagram, so that the resulting color contrast is strong.
When both GST layers are in the amorphous phase, the structure appears red, while in the crystalline phase, it appears turquoise. 
We found that, as the incident angle increased, the reflection resonances shifted toward shorter wavelengths. This shift was very small for incident angles up to $15^\circ$, but became more pronounced beyond this threshold.

We also applied the memetic optimization algorithm to optimize multilayer structures for producing four maximally distinct colors. In this case, our eight-layer design features two thin GST layers. The top GST layer lies beneath layers of MgF$_2$ and SiC, while the bottom GST layer is placed above layers of MgF$_2$ and SiC on top of the silver substrate. Dielectric layers of HfO$_2$ and Al$_2$O$_3$ are used to separate the two GST layers.
This structure produces blue, yellow, red, and cyan colors in the CC, AA, CA, and AC states, respectively. 
The minimum pairwise color distance achieved is $\sim$0.27, surpassing the contrast performance of previously reported structures. As with the two-color design, the reflection spectra in all four states remained nearly identical to those at normal incidence for incident angles up to $15^\circ$. We also found that increasing the number of layers in both the two-color and four-color designs does not significantly enhance the distinctiveness of the generated colors.

As final remarks, the proposed multilayer structures can be fabricated using standard thin-film deposition techniques, such as sputtering, thermal evaporation, or electron-beam evaporation. A resistive layer, such as Pd/NiCr, can be integrated underneath the structure to enable switching of the PCM layers. This layer provides Joule heating without direct contact with the PCM layers \cite{jafari_2019_1}. Unlike direct heating methods, where the current flows through the PCM itself, in this indirect approach, the current flows only through the resistive layer, thereby improving device reliability and extending the number of switching cycles \cite{sreekanth2018_1_37, wang2014_1_38}. By adjusting the amplitude and duration of short electrical pulses applied to the resistive layer, the temperature can be raised to selectively switch one of the PCM layers while leaving the other unchanged \cite{jafari_2019_1}. 
In addition to their color tunability, the proposed multilayer structures eliminate the need for subwavelength lithography, often required in metasurface-based color filters, making them particularly suitable for large-scale applications.
Our findings could pave the way for a new class of single-cell multicolor pixels with no power consumption required to retain each color, making them particularly appealing for low refresh rate displays. 

\section*{Funding}
National Science Foundation (2150491).

\section*{Disclosures}
The authors declare no conflicts of interest.

\section*{Data Availability}
Data underlying the results presented in this paper are not publicly available at this time but may be obtained from the authors upon reasonable request.

\bibliography{reference}

\end{document}